\documentclass[letterpaper,floatfix,nobalancelastpage,twocolumn,superscriptaddress]{revtex4}
\usepackage{overpic}
\usepackage{dcolumn}
\usepackage{bm}
\usepackage{amsmath, amsfonts,amssymb}
\usepackage{bm,braket}
\usepackage{color}
\usepackage{float}
\usepackage{graphicx}
\usepackage{subfigure}
\usepackage{soul}
\usepackage{comment}
\usepackage[utf8x]{inputenc}

\usepackage{sidecap}
\newcommand{\rep}[1]{{\bf\color{blue}#1}}

\newcommand{\rem}[1]{}

\definecolor{violet}{rgb}{0.58, 0.0, 0.83}

\newcommand{\mkaddcomment}[1]{{\bf {\color{violet}{[MK: #1]}}}}

\begin{document}

\title{Floquet engineering of lattice structure and dimensionality in twisted moir\'e heterobilayers}
\author{Rong-Chun Ge}
\author{Michael Kolodrubetz}
\affiliation{Department of Physics, The University of Texas at Dallas, Richardson, Texas 75080, USA}
\date{\today}

\begin{abstract}
We present an experimental proposal to tune the effective lattice structure in twisted transition metal dichalcogenide (TMD) heterobilayers with time-periodic Floquet drive. We show that elliptically polarized light with sub-terahertz frequencies $\hbar\omega\sim 1$ meV and moderate electric fields $E\sim0.2$~MV/cm allows tuning between the native triangular lattice and a square lattice, while linearly polarized light enables dimensional reduction to a quasi-one-dimensional geometry. Without drive, these twisted TMDs simulate the single band Fermi-Hubbard model; we show that this approximation still holds in the presence of drive. This control opens the door to explore a rich variety of correlated phases of matter, such as spin liquids and d-wave superconductivity.
\end{abstract}
\maketitle

The ability to stack monolayers of two dimensional materials with small twist angles between layers has opened up a new and exciting platform for strongly correlated physics. 
When two such periodic lattices are stacked together, a moir\'e pattern emerges with a much larger lattice constant than either individual lattice. Numerical calculations showed that flat bands emerge at specific ``magic'' twist angles between the layers~\cite{M1,M2,M3}, and recent experiments supported these findings by demonstrating singularities of the density of states~\cite{DOS1,DOS2,DOS3,DOS4,DOS5,DOS6} and experimental dispersion relations consistent with the theory~\cite{Band1}. Even away from these magic angles, the kinetic energy is suppressed compared to the interaction energy, with the relative strength of on-site interactions under experimental control via the twist angle. A wide range of strongly correlated phases~\cite{CORR,CORR1,CORR2,CORR3} including unconventional superconductivity~\cite{SUPER} and Wigner crystals~\cite{Wigner} have been discovered in these systems, and a wider range of phases is expected as new materials, structures, and control techniques are explored.

Beyond these equilibrium properties, two-dimensional moir\'e materials have excellent potential for non-equilibrium engineering of their Hamiltonian, notably through time-periodic Floquet control via lasers. Floquet engineering in atomic and molecular systems has been a major experimental triumph of the past decade \cite{Fana0,Fana1,Fana2,DDcou1,DDcou2,DDcou3,POD}, achieving successes such as extremely large effective magnetic fields for charge neutral particles~\cite{Raman,Raman1,Raman2,Mag1,Mag2} and novel topologically protected phases of matter ~\cite{FT1,FT2,FT3,FT4,FT5,FT6,FT7}. Floquet engineering has had limited success in electronic systems due to the relatively small lattice constant compared to optical lattices used for ultracold atoms, which translates into very large electric field strength of order tens of MV/cm to achieve useful Floquet control on the lattice scale \cite{FTSolid}.

\begin{figure}[b]
\centering
\includegraphics[width=.8\columnwidth]{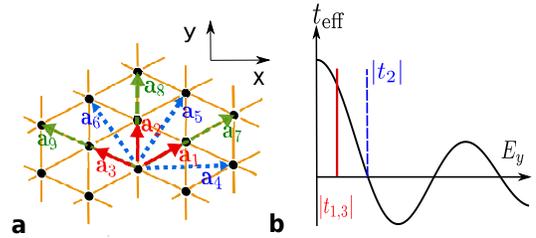}
\caption{(a) Illustration of triangular moir\'e lattice structure formed by a TTMDH, which we modify via Floquet drive. 
 The potential landscape admits beyond nearest neighbor hopping. Leading tunneling terms are illustrated here, where lattice vectors $a_i$ correspond to tunneling coefficients $t_i$. (b) We propose to use electromagnetic driving with in plane polarization to control the effective geometry of the moir\'e lattice. For $y$ polarization, the nearest neighbor hoppings can be tuned from triangular lattice to square lattice by adjusting the field amplitude, a phenomenon known as coherent destribution of tunneling. 
}
\label{f0}
\end{figure}

By contrast, moir\'e lattices resulting from these 2D heterstructures have lattice spacings of order 100 times the original 2D materials, which makes Floquet engineering much more practical. In this work, we apply ideas of Floquet control to engineer geometry and dimensionality of moir\'e lattice structures. Specifically, we consider twisted transition metal dichalcogenide heterobilayers (TTMDH), which have been shown theoretically~\cite{2DHUB} and experimentally~\cite{2DHUB1} to realize an isotropic single band Hubbard model on a triangular lattice. We show that the geometry of the Hubbard model can be tuned from a triangular to square lattice using either a single elliptically polarized beam or two linearly polarized beams at frequencies $\omega$ and $2\omega$.  By working at frequencies low compared to the bandwidth, we show that a relatively moderate electric field of order 0.2~MV/cm is required, well within the range of current experiments. Control of lattice geometry will allow direct exploration of a wide variety of theoretically predicted phases, such as superconductivity~\cite{square0}, antiferromagnetism, strange metallic phases~\cite{square,square1}, and spin liquids~\cite{tri0,tri1,tri2,tri3}. We note that a pair of recent papers suggested similar Floquet control of the moir\'e lattice structure of twisted graphene bilayers to realize Chern insulators~\cite{FM1,FM2}. Our work differs from these in two main ways: 1) our goal is to modify the lattice structure in order to achieving different strongly correlated phases of matter rather than weakly correlated topological phases of matter and 2) our working point is at a much lower frequency and electric field amplitude, which may render it more experimentally feasible.
   
\textbf{\textit{Model}} -- We will study the continuum model of electrons on the moir\'e lattice at small twist angle \cite{M2}, for which Ref.~\cite{2DHUB} showed that a single band Hubbard model on triangular lattice could be obtained with TTMDHs, a prediction later confirmed experimentally~\cite{2DHUB1}.
Following that proposal, we will work with WSe$_2$/MoSe$_2$ for which the highest valence band of WSe$_2$ lies inside the gap of MoSe$_2$. The moir\'e band Hamiltonian (valence band) is given by~\cite{M2,2DHUB}
\begin{align}
H = -\frac{{\bf p}^2}{2m_{\rm eff}}+\Delta({\bf r})
\label{CM}
\end{align}
with potential landscape $\Delta({\bf r}) = \sum_jV({\bf b}_j)\exp(i{\bf b}_j\cdot {\bf r})$ ($j=1,2,3$) satisfying $C_3$ rotational symmetry, where ${\bf b}_1 = 4\pi/(\sqrt{3}a_M) \hat{x}$ with $a_M$ the moir\'e lattice constant, ${\bf b}_{j+1~\mathrm{mod}~3} = \mathcal{R}_{2\pi/3}{\bf b}_j$, $V({\bf b}_j) = V^*(-{\bf b}_j)$, and $V({\bf b}_j)=V({\bf b}_{j+1~\mathrm{mod}~3})$. The parameters are chosen to be $V=6.6$ meV, $\phi=-94^\circ$, and $m_{\rm eff}=0.35 m_e$, where $m_e$ is the free electron mass, based on a fit to {\it ab initio} numerics in Ref. \cite{2DHUB}.

For numerical calculations, we choose a twist angle $2^\circ$, for which the moir\'e lattice constant is $a_M = 9.4~$nm. 
The highest occupied valence band is well separated from the rest of higher energy states, and it is well described by a tight-binding model with up to third neighbor tunneling
\begin{align}
H_t = \sum_{ij\sigma} (t_{j}c_{{\bf R}_i,\sigma}^\dagger c_{{\bf R}_i+{\bf a}_j,\sigma}+h.c.).
\label{TB}
\end{align}
Here ${\bf a}_i$ are the lattice vectors shown in Fig.~\ref{f0}(a) with $\left|{\bf a}_{1,2,3}\right| = a_M$,
 ${\bf R}_i$ are the sites of the triangular lattice, and $\sigma$ is the spin. The tunneling coefficient along ${\bf a}_i$ is $t_i$. By symmetry, we have $t_{1/2/3} = J$ for nearest neighbors, $t_{4/5/6} = \alpha J$ for next nearest neighbors, and $t_{7/8/9} = \beta J$ for next next nearest neighbors in the absence of drive. Fitting the band structure, we find $J=1.34~$meV, $\alpha=-0.14$, and $\beta=-0.08$, as in~\cite{2DHUB}. 

We will consider Floquet control of the band structure via time-periodic electric field due to an incident beam. The minimal coupling Hamiltonian is given by
\begin{align}
H' = -\frac{({\bf p}-e{\bf A(t)})^2}{2m_{\rm eff}}+\Delta({\bf r}).
\label{ConT}
\end{align}
where the drive is polarized in the plane of the TTMDH: ${\bf A}({\bf r},t) = [A_xf(\omega t),A_y g(\omega t)]$ for $2\pi$-periodic functions $f$ and $g$. The intuition for Floquet control of lattice geometry comes from the phenomenon of coherent destruction of tunneling~\cite{Ctrapping1,Ctrapping2}, 
 in which the drive induces destructive interference for tunneling parallel to the polarization. For large driving frequencies, the hoppings are effectively modified by the drive and can be approximated through the leading order term of a high-frequency expansion (HFE) such as the Magnus expansion \cite{FQE1,FQE2}. 
  In the case of linear $y$-polarization, the HFE predicts that nearest neighbor hopping in the vertical direction is turned off ($t_2^{\rm HFE} = 0$) for $e A_y a_M / \hbar \approx 2.4$, resulting in a nearest neighbor lattice geometry equivalent to a square lattice. In conventional electronic systems, this requires inacessably strong fields due to the small lattice constant. For the moir\'e superlattice, lower frequencies and drive amplitudes are possible, but they render the HFE suspect due to resonances with higher bands. We will now show that, nevertheless, such a high frequency expansion technique works surprisingly well even for relatively low frequencies ($\hbar \omega \sim 1~$meV), allowing meaningful Floquet engineering of band structures in TTMDHs.

\textbf{\textit{Results}} -- In order to show that Floquet engineering can be achieved at low drive frequency and amplitude, we consider an elliptically polarized beam with vector potential given by \footnote{We will not consider the effect of phonons here, although we note that they may have a notable effect as shown in Ref.~\cite{FM2} for a related model.}
\begin{equation}
{\bf A}(t)=[A_x\cos(\omega t+\phi),A_y \sin(\omega t)].
\end{equation}
Applying this drive to the hopping model and considering the high frequency expansion, one readily finds expressions for the effective tunneling constants, which are shown in the Supplement~\cite{SM}. 
For linear $y$-polarization, $t_2$ is given at leading order by
\begin{equation*}
t_2^{\rm HFE} = t_2 J_0(e A_y a_M / \hbar).
\end{equation*}
Coherent destruction of tunneling is achieved by working at $A_y$ values such that $e A_y a_M / \hbar$ is at a zero of the Bessel function. Meanwhile, we have $t_1^{\rm HFE}=t_3^{\rm HFE}\neq 0$~\cite{SM}, resulting in a square lattice geometry for nearest neighbor couplings. However, the longer range hopping causes an issue. For instance, $t_4$ remains unchanged by the drive, resulting in an anisotropic lattice configuration rather than the desired square lattice. 

\begin{figure}[t]
\centering
\includegraphics[width=\columnwidth]{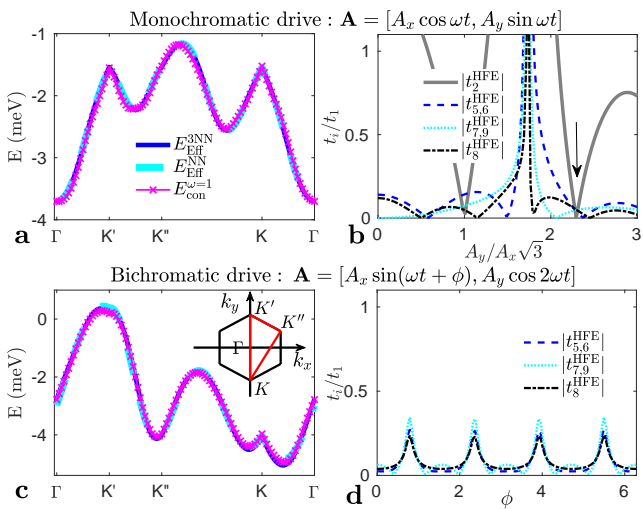}
\caption{Floquet engineering of the TTMDH using monochromatic (a,b) or bichromatic (c,d) driving with $\hbar \omega = 1$ meV. (a) Effective band structure for monochromatic driving with $\sqrt{3}e A_x a_M/\hbar = 2.4$ and $e A_y a_M/\hbar = 5.52$, showing that the full Floquet treatment of the continuum model (``con,'' pink) matches well with the high-frequency expansion of the third nearest-neighbor hopping model (``NNN,'' blue) or just nearest neighbors (``NN,'' red). The cut taken through the first Brillouin zone is shown in the inset to (c). (b) Amplitude dependence of the effective hoppings within the high-frequency expansion. (c) Effective band structure for bichromatic driving with $\sqrt{3}e A_x a_M/\hbar = 5.52$ and $e A_y a_M/\hbar = 2.4$ and relative phase $\phi=0$. 
The dispersion is again well-approximated by a square lattice. Differences from (a) come from an overall rescaling of the tunnel couplings and non-zero phase on the effective hoppings, resulting in a shift of the band minimum. (d) $\phi$-dependence of the tunnel couplings, showing that the square lattice is relatively stable, i.e., $t_{i > 3}$ remain small. By symmetry, $\left| t_1^{\rm HFE} \right| = \left| t_3^{\rm HFE} \right|$ for both drives.
}
\label{f1}
\end{figure}

To suppress $t_4$ and bring the effective Hamiltonian closer to that of a square lattice, we use elliptical polarization. As seen in Fig.~\ref{f1}(b), the drive parameters $\hbar \omega=1~$meV, $eA_ya_M/\hbar = 5.52$, and $\sqrt{3}e A_x a_M/\hbar = 2.4$ -- corresponding to $|E|\sim2.6\times 10^5~$V/cm -- yield simultaneous zeroing of $t_2$ and $t_4$ within the HFE.
Furthermore, further neighbor terms are made sufficiently small compared to the renormalized $t_{1,3}^{\rm HFE}$ that the system may be well-approximated by an isotropic square lattice with predominately nearest neighbor hopping. Note that, by interpolating between the undriven case and the parameters shown here, one can tune from a triangular lattice to square lattice, introducing controllable anisotropy for realizing strongly correlated phases of matter; we will comment on this later.

The main question, then, is whether these predictions from the HFE hold up despite the low frequencies chosen, for which resonances with higher energy states within the continuum model could lead to heating or modification of the band structure. To show that $\hbar \omega \sim 1$ meV works, we directly solve the Floquet quasienergies of the driven continuum model (Eq.~\ref{ConT}). Unlike the effective Hamiltonian within the HFE, these quasienergies are only well-defined modulo units of the photon energy, $\hbar \omega$. Therefore, to compare the models, we project the Floquet evolution operator to the highest occupied valence band in order to determine the relevant Floquet state (and it's leakage from the valence band) and choose the branch such that quasienergy is a continuous function of momentum~\cite{FA1}. 
We find that $\hbar \omega=1$ meV is a sweet spot in which the HFE provides accurate predictions while mixing to higher energy states remains negligible; data for other frequencies is shown in the supplement for comparison. 


It's worth noting that there is significant freedom in the Floquet control technique. For instance, a similar lattice tuning can be achieved with bichromatic drive with $x$ and $y$ components driven at at frequencies $\omega$ and $2\omega$ respectively. 
 As shown in  Fig.~\ref{f1}(b), such bichromatic driving has the further advantage that the tunneling coefficients only weakly depend on the phase difference $\phi$ between the drives, which be useful experimentally due to difficulties synchronizing these phases.

Finally, we note that one can instead use linear polarization to reduce the dimensionality by tuning the polarization to be perpendicular to one of the nearest neighbor hopping directions (e.g., $b_3$). As shown in the supplement \cite{SM}, hopping along this direction $t_3$ is unmodified, while the other nearest neighbor tunnelings can be turned off by tuning to a zero of the Bessel function. Similar to the square lattice, this protocol can be adapted slightly to decrease longer range tunneling, enabling the study of 1D systems with this platform.

\textbf{\textit{Experiments}} -- The drive protocols we introduce here involve both frequencies and intensities that are accessible by modern experimental techniques~\cite{FControl1,FControl2}. 
 The electric field is still high enough that these drives will initially be accessible via pulsed sources, such that the experiments we envision would involve pump-probe detection of the novel strongly correlated physics. However, we note that the development of improved sources in this terahertz frequency range remains an ongoing topic of experimental interest, suggesting that eventually a continuous wave drive or longer pulses may be available.

These driven systems will be a crucial platform to search for new strongly correlated phases of matter due to the ability to tune hopping strength and geometry. As predicted in Ref.~\cite{2DHUB}, TTMDH are well-approximated by the Fermi-Hubbard model, in which Coulomb interactions only enter as an on-site term $\sum_{i\sigma}U n_{{\bf R}_i\sigma} n_{{\bf R}_i\bar\sigma}$. Since interactions are not modified by the drive considered here~\cite{SWTForperiodic}, 
we therefore predict that experiments will be able to demostrate such a Fermi-Hubbard model on a tunable lattice geometry. The value of $U$ is tunable by twist angle or electromagnetic environment. It is estimated to be of order $20$ to $30$ meV in current experiments~\cite{2DHUB,2DHUB1}. 

\begin{figure}[t]
\centering
\includegraphics[trim=.0cm 0.0cm 0.0cm .0cm, clip=true,width=.69\columnwidth]{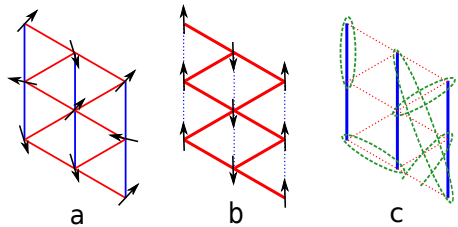}
\caption{Predicted phases accessible with Floquet control at half-filling. (a) 120$^\circ$ antiferromagnetic for isotropic triangular lattice. (b) Neel state on square lattice. (c) Spin liquid with anisotripic triangular lattice $|t_2^{\rm HFE}/t_{1/3}^{\rm HFE}|>1$. $t_2^{\rm HFE}$ and $t_{1/3}^{\rm HFE}$ are represented by the blue and red respectively. 
}
\label{f2}
\end{figure}

The Fermi-Hubbard model is widely considered to be a minimal model for high-temperature superconductivity and serves as a parent Hamiltonian for a cornucopia of fascinating strongly correlated phases of matter. Accessing it in these two-dimensional materials gives direct control over filling via gating, so the model can be studied either at half-filling or in the presence of doping. At half-filling, for $U\gg t$, we recover the nearest-neighbor Heisenberg model whose exchange couplings $J_i =  2|t^{\rm HFE}_{i}|^2/U$ depend on the lattice geometry. Even within the Heisenberg model, interesting phases have been predicted as one tunes anisotropy to go between triangular and square lattices. The idea is illustrated in Fig.~\ref{f2}. For the triangular or square lattice one gets an antiferromagnetic state, with 120$^\circ$ or Neel order respectively. In between these limits, neither ordering is commensurate, and mean field calculations have suggested the possibility of a non-trivial magnetically disordered state -- the quantum spin liquid \cite{SP1,SP2}. Furthermore, quantum Monte Carlo has predicted that if one instead increases the anisotropy via $t_2 > t_{1,3}$, two other spin liquid phases become possible~\cite{Monte}. This is even easier for our drive protocols, as it uses a lower drive strength in both the $x$ and $y$ directions. These predictions have never been verified due to lack of tunability in existing (quasi)-2D materials. Our platform, therefore, provides a useful route to test these predictions and look for new spin liquid phases. 

The potential for other strongly correlated phases becomes even greater as one moves away from this Heisenberg limit. Density matrix renormalization group calculations have examined the rich phase diagram for finite on-site interaction $U$. For square lattices, this is the regime where one finds $d$-wave superconductivity and stripe order away from half filling, as seen in cuprate superconductors~\cite{square,square1}. For triangular lattices at half filling, spin liquid phases have been predicted \cite{tri1,tri2} which seem to survive finite anisotropy \cite{tri3}. As other factors are added, such as tuning away from half filling, adding lattice anisotropy, and considering longer range interactions, a zoo of phases has been predicted similar to those in twisted graphene structures, including correlated insulators~\cite{CORR3} 
 and generalized Wigner crystals \cite{Wigner}. Despite the fact that the anisotropic Fermi-Hubbard model has been well studied through a variety of theoretical techniques, there are still a number of discrepancies between the predictions. By exploiting our experimental proposal, TTMDH may serve as a testbed for these theoretical predictions.

\textbf{\textit{Conclusion}} -- We have presented an experimental proposal to engineer the geometry of a moir\'e lattice configuration using Floquet drive, tuning the native triangular lattice to a square lattice or dimensionally reducing it to 1D. By using a sub-bandwidth frequency of $\hbar \omega = 1~$meV, the requisite intensity is much weaker than previous suggestions for Floquet engineering in the solid state, bringing it to an experimentally accessible regime. Interactions yield a Fermi-Hubbard model with tunable relative interaction strength, which serves as a template for many interesting correlated phases of matter.

The basic principles here should be accessible in other 2D moir\'e structures, such as twisted graphene, enabling further control over the novel phases of matter under investigation in those systems. Other important effects, such as phononic contributions \cite{FM1}, may qualitatively modify the structure of these phases of matter, which is an important topic for further investigation. While future experiments may be able to realize these phases in a late-time steady state with continuous wave drive, current experiments are more likely to use pump-probe methods which have been successfully demonstrated for, e.g., Floquet engineering of topology in graphene. The non-equilibrium behavior of these strongly correlated phases of matter in such experiments is another fascinating challenge, which may yield non-trivial behavior through non-equilibrium effects such as the Kibble-Zurek mechanism~\cite{KZ1,KZ2,KZ3,KZ4}.



\textbf{\textit{Acknowledgments}} --
This work was performed with support from the National Science Foundation through award number DMR-1945529 and the Welch Foundation through award number AT-2036-20200401. We used the computational resources of the Lonestar 5 cluster operated by the Texas Advanced Computing Center at the University of Texas at Austin and the Ganymede and Topo clusters operated by the University of Texas at Dallas' Cyberinfrastructure \& Research Services Department

\section{Supplementary material}
\label{ap}
\subsection{Renormalized tunneling coefficients}

We obtain the renormalized tunneling coefficients by a conventional Floquet high-frequency expansion (HFE), as described in \cite{FQE1}. This begins by inserting the vector potential into the tunneling Hamiltonian via a Peierl's substitution:
\begin{equation*}
H_t = \sum_{ij\sigma} (t_{j}c_{{\bf R}_i,\sigma}^\dagger c_{{\bf R}_i+{\bf a}_j,\sigma} \exp\left[ ie{\bf A}(t) \cdot {\bf a}_j / \hbar \right]+h.c.).
\end{equation*}
where the lattice vectors are given by ${\bf a_{1/3}} = (\pm\frac{\sqrt{3}}{2},\frac12){a}_M=\frac12{\bf a_{7/9}}$,  ${\bf a_2} = (0,1){a}_M=\frac12{\bf a_8}$,
${\bf a_4} = (\sqrt{3},0){a}_M$, and ${\bf a_{5/6}} = (\pm\frac{\sqrt{3}}{2},\frac32){a}_M$. The exponential is expanded as 
\begin{equation*}
\exp\left[ ie A_x(t) a_{j,x} / \hbar \right] \exp\left[ ie A_y(t) a_{j,y}  / \hbar \right],
\end{equation*}
where for all of the protocols we consider, $A_{x,y} (t)$ are simple sinusoidal functions. For illustration, we consider the monochromatic driving scheme, for which ${\bf A} = [A_x\cos(\omega t+\phi),A_y\sin(\omega t)]$. Then each term of the exponential is expanded by the Jacobi-Anger formula, e.g.,
\begin{equation*}
\exp\left[ ie A_x(t) a_{j,x}  / \hbar \right] = \sum_{n=-\infty}^{\infty} i^n J_n\left( e A_x a_{j,x}  / \hbar \right) e^{i n (\omega t + \phi)}.
\end{equation*}
Finally, the resulting Hamiltonian is time-averaged, giving the leading (zeroth order) term in the high-frequency expansion.

For the monochromatic driving scheme with ${\bf A} = [A_x\cos(\omega t+\phi),A_y\sin(\omega t)]$, the tunneling coefficients of the effective Hamiltonian at the first order approximation with are 
\begin{align}
t_{1}^{\rm HFE}=&\sum_nJ_n(eA_ya_M/2\hbar)J_{n}(eA_x\sqrt{3}a_M/2\hbar)i^{n}e^{-in\phi}t_1\nonumber\\
t_2^{\rm HFE}=&J_0(eA_ya_M/\hbar)t_1\nonumber\\
t_3^{\rm HFE}=&\sum_nJ_n(eA_ya_M/2\hbar)J_{n}(-eA_x\sqrt{3}a_M/2\hbar)i^{n}e^{-in\phi}t_1\nonumber\\
t_5^{\rm HFE}=&\sum_nJ_n(3eA_ya_M/2\hbar)J_{n}(eA_x\sqrt{3}a_M/2\hbar)i^{n}e^{-in\phi}t_4\nonumber\\
t_6^{\rm HFE}=&\sum_nJ_n(3eA_ya_M/2\hbar)J_{n}(-eA_x\sqrt{3}a_M/2\hbar)i^{n}e^{-in\phi}t_4\nonumber\\
t_7^{\rm HFE}=&\sum_nJ_n(eA_ya_M/\hbar)J_{n}(eA_x\sqrt{3}a_M/\hbar)i^{n}e^{-in\phi}t_7\nonumber\\
t_8^{\rm HFE}=&J_0(2eA_ya_M/\hbar)t_7\nonumber\\
t_9^{\rm HFE}=&\sum_nJ_n(eA_ya_M/\hbar)J_{n}(-eA_x\sqrt{3}a_M/\hbar)i^{n}e^{-in\phi}t_7\nonumber\\
\end{align}
By choosing $eA_ya_M/\hbar = 5.52$ (second zero point of $J_0(x)$) and $\sqrt{3} eA_xa_M/\hbar = 2.4$ (first zero point of $J_0(x)$), we have $t_2^{\rm HFE}=t_4^{\rm HFE}=0$.

For bichromatic driving with ${\bf A} = [A_x\sin(\omega t+\phi),A_y\cos(2\omega t)]$, the tunneling coefficients of the effective Hamiltonian are 
\begin{align}
t_1^{\rm HFE}=&\sum_nJ_n(eA_ya_M/2\hbar)J_{2n}(eA_x\sqrt{3}a_M/2\hbar)i^{n}e^{-2in\phi}t_1\nonumber\\
t_2^{\rm HFE}=&J_0(eA_ya_M/\hbar)t_1\nonumber\\
t_3^{\rm HFE}=&\sum_nJ_n(eA_ya_M/2\hbar)J_{2n}(-eA_x\sqrt{3}a_M/2\hbar)i^{n}e^{-2in\phi}t_1\nonumber\\
t_5^{\rm HFE}=&\sum_nJ_n(3eA_ya_M/2\hbar)J_{2n}(eA_x\sqrt{3}a_M/2\hbar)i^{n}e^{-2in\phi}t_4\nonumber\\
t_6^{\rm HFE}=&\sum_nJ_n(3eA_ya_M/2\hbar)J_{2n}(-eA_x\sqrt{3}a_M/2\hbar)i^{n}e^{-2in\phi}t_4\nonumber\\
t_7^{\rm HFE}=&\sum_nJ_n(eA_ya_M/\hbar)J_{2n}(eA_x\sqrt{3}a_M/\hbar)i^{n}e^{-2in\phi}t_7\nonumber\\
t_8^{\rm HFE}=&J_0(2eA_ya_M/\hbar)t_7\nonumber\\
t_9^{\rm HFE}=&\sum_nJ_n(eA_ya_M/\hbar)J_{2n}(-eA_x\sqrt{3}a_M/\hbar)i^{n}e^{-2in\phi}t_7\nonumber\\
\end{align}
Now, by choosing $eA_ya_M/\hbar = 2.4$ (first zero point of $J_0(x)$) and $eA_x\sqrt{3}a_M/\hbar = 5.52$ (second zero point of $J_0(x)$), we again achieve $t_2^{\rm HFE}=t_4^{\rm HFE}=0$.

\begin{figure}
\centering
\includegraphics[width=\columnwidth]{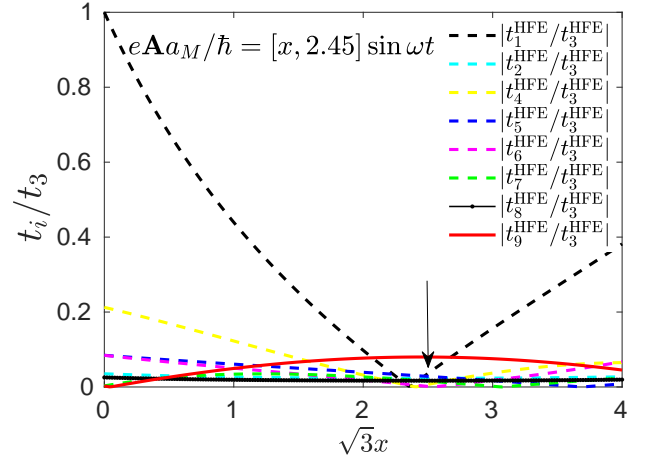}
\caption{Engineering of one-dimensional lattice with linearly polarized drive. By tuning to $e{\bf A}a_M/\hbar = [2.47/\sqrt{3},2.45]\sin(\omega t)$ (arrow), all hopping terms are strongly suppressed besides $t_3$ and $t_9$.}
\label{SM3}
\end{figure}

Finally, as mentioned in the main text, it is possible to get decoupled one dimensional systems using a linearly polarized drive, which is a subset of the monochromatic driving shown above. The tunneling coefficients from the HFE are plotted in Fig.~\ref{SM3} near, but not quite at, a zero of Bessel function. The data shows that for $e{\bf A}a_M/\hbar = [2.47/\sqrt{3},2.45]\sin(\omega t)$, we have $t_9^{\rm HFE}/t_3^{\rm HFE}\approx 0.08$ and $t_j^{\rm HFE}/t_3^{\rm HFE}<0.03$, which means that tunneling is suppressed except in the ${\bf a}_3$ direction, resulting in an effective one dimensional system.

\subsection{Floquet analysis}
In this subsection, we give more details about the Floquet analysis of the continuous model~\cite{FA1}.

For the time periodic but spatially uniform external field ${\bf A}(t)$ our system is described by
\begin{align}
i\hbar\frac{\partial\Psi(x,y,t)}{\partial t} &= H^\prime \Psi(x,y,t) \nonumber \\
H^\prime \equiv & -\frac{({\bf p}-e{\bf A(t)})^2}{2m_{\rm eff}}+\Delta({\bf r})\label{eq:Shrodinger}.
\end{align}
Since, ${\bf A}(t+T) = {\bf A}(t)$, the behavior of the system in the long time limit is given by the Floquet operator,
\begin{align}
U(T) = \mathcal{T}\exp(-\frac{i}{\hbar}\int_0^T H'(t)dt).
\end{align}
The Hamiltonian is periodic at any time, as a result the Bloch wavevector ${\bf k}_B$ is a good quantum number; so the Floquet operator could be simplified as $U(T) = \bigoplus_{{\bf k}_B} U_{{\bf k}_B}(T)$. The quasienergy which is well defined up to a multiplication of $2\pi$ is given by $\varepsilon_{\rm quasi}({\bf k}_B) = i\hbar\log(\mathcal{E}[U_{{\bf k}_B}(T)])/T$.

Numerically we use the plane wave expansion method: starting from $t=0$, we obtain the spectra of the lowest band of $H'(t=0)$, $\varepsilon_0({\bf k}_B),\,|\psi_{0}({\bf k}_B)\rangle$, and the projector operator $\mathcal{P}_{{\bf k}_B} = |\psi_0({\bf k}_B)\rangle\langle\psi_0({\bf k}_B)|$. The complete evolution operator $U_{{\bf k}_B}$ is obtained as $\langle {\bf k}_i |U_{{\bf k}_B}|{\bf k}_j\rangle = \langle {\bf k}_i|\mathcal{T}\exp(-\frac{i}{\hbar}\int_0^TH'(t)dt)|{\bf k}_j\rangle\sum_{n_1,n_2}\delta_{{\bf k}_B,n_1{\bf G}_1+n_2{\bf G}_2+{\bf k}_j}$ where $n_i$ are integers, $i=1,2$, and ${\bf G}_i$,  are the reciprocal lattice constants. The full operator is projected into the Hilbert space spanned by the first band, $U^P_{{\bf k}_B} = \mathcal{P}_{{\bf k}_B}U_{{\bf k}_B}\mathcal{P}_{{\bf k}_B}$. If the projected space is isolated from the rest of the Hilbert space, $U_{{\bf k}_B}^P$ will be unitary, with norm one eigenvalues. Otherwise, the modulus of the eigenvalues will deviate from the unit; which for initial population in the spanned space means the decay of the population, and is described by a nonzero imaginary part ${\rm Im}[\varepsilon_{\rm quasi}]$. 
\begin{figure}
\centering
\includegraphics[width=\columnwidth]{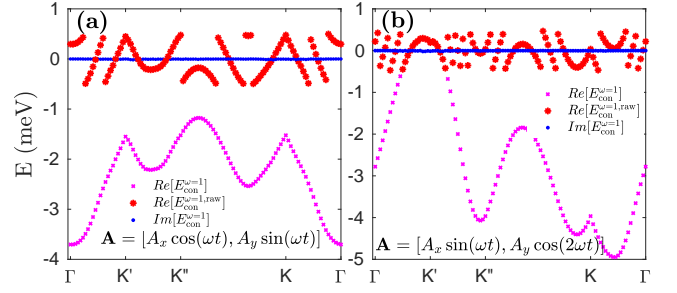}
\caption{(Quasi-)Energy spectra of monochromatic (a) or bichromatic (b) driving with $\hbar \omega = 1$ meV. (a) Effective band structure for monochromatic driving with $\sqrt{3}e A_x a_M/\hbar = 2.4$ and $e A_y a_M/\hbar = 5.52$. (b) Effective band structure for bichromatic driving with $\sqrt{3}e A_x a_M/\hbar = 5.52$ and $e A_y a_M/\hbar = 2.4$ and relative phase $\phi=0$. 
Red stars are the raw data of real part of the energy, ${\rm Re}[\varepsilon(k)]$, blue dots are the imaginary part, ${\rm Im}[\varepsilon(k)]$, and the magenta crosses are the clear data according to time binding calculation shown in the main text. 
}
\label{SM1}
\end{figure}

As is shown clearly in Fig.~\ref{SM1} the imaginary part of the spectra is negligible for both the monochromatic and bichromatic driving. So we can effective separate the single band from the rest of the system.

\begin{figure}
\centering
\includegraphics[width=\columnwidth]{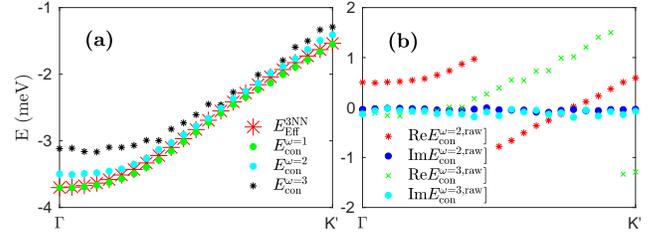}
\caption{(Quasi-)Energy spectra of monochromatic driving for $\sqrt{3}e A_x a_M/\hbar = 2.4$ and $e A_y a_M/\hbar = 5.52$
 . (a) Effective band structure for $\hbar \omega=1,2,3~$meV (green/cyan,black dots); the red stars are high-frequency expansion of the third nearest-neighbor hopping model. (b) Raw data for the quasi-energy of: red and green stars are the real part of the quasienergies for $\hbar \omega=2/3~$meV respectively; and the blue and cyan dots are the imaginary part of the quasienergies for $\hbar \omega=2,3~$meV.    
}
\label{SM2}
\end{figure}

From the our analysis in the main text, as the driving frequency increases, the group band gets coupled to higher bands gradually; and this is indicated by the finite imaginary part of the dipersion relation as is shown in Fig.~\ref{SM2}(b). As a result a single band model will not be possible, and this is also demonstrate from the deviation of calculation from the continuous model from that of the tight-binding model in Fig.~\ref{SM2}(a).

\end{document}